\documentclass[prx,twocolumn,superscriptaddress,showpacs,amsmath,amstex,amssymb,citeautoscript,nofootinbib]{revtex4-2}
\usepackage{dcolumn}% Align table columns on decimal point
\usepackage{bm}% bold math
\usepackage[T1]{fontenc}
\usepackage{braket}
\usepackage{graphicx}
\usepackage{xcolor}
\usepackage{lipsum}
\usepackage[normalem]{ulem}
\usepackage{color}
\usepackage[percent]{overpic}
\usepackage{soul} 
\usepackage{wasysym}
\usepackage{dsfont}
\usepackage{float}
\usepackage[mathscr]{euscript}
\usepackage{csquotes}
\usepackage{appendix}
\usepackage[bottom]{footmisc}
\usepackage{verbatim}
\usepackage{multirow}
\usepackage{cleveref}
\crefformat{equation}{Eq.~#2#1#3}
\crefformat{figure}{Fig.~#2#1#3}

\begin{document}

\title{Catching Bethe phantoms and quantum many-body scars:\\
Long-lived spin-helix states in Heisenberg magnets}

\author{Paul Niklas Jepsen}
\affiliation{Department of Physics, Massachusetts Institute of Technology, Cambridge, MA 02139, USA}
\affiliation{Research Laboratory of Electronics, Massachusetts Institute of Technology, Cambridge, MA 02139, USA}
\affiliation{MIT-Harvard Center for Ultracold Atoms, Cambridge, MA, USA}

\author{Yoo Kyung Lee}
\affiliation{Department of Physics, Massachusetts Institute of Technology, Cambridge, MA 02139, USA}
\affiliation{Research Laboratory of Electronics, Massachusetts Institute of Technology, Cambridge, MA 02139, USA}
\affiliation{MIT-Harvard Center for Ultracold Atoms, Cambridge, MA, USA}

\author{Hanzhen Lin}
\affiliation{Department of Physics, Massachusetts Institute of Technology, Cambridge, MA 02139, USA}
\affiliation{Research Laboratory of Electronics, Massachusetts Institute of Technology, Cambridge, MA 02139, USA}
\affiliation{MIT-Harvard Center for Ultracold Atoms, Cambridge, MA, USA}

\author{Ivana Dimitrova}
\affiliation{MIT-Harvard Center for Ultracold Atoms, Cambridge, MA, USA}
\affiliation{Department of Physics, Harvard University, Cambridge, MA 02138, USA}

\author{Yair Margalit}
\affiliation{Department of Physics, Massachusetts Institute of Technology, Cambridge, MA 02139, USA}
\affiliation{Research Laboratory of Electronics, Massachusetts Institute of Technology, Cambridge, MA 02139, USA}
\affiliation{MIT-Harvard Center for Ultracold Atoms, Cambridge, MA, USA}

\author{Wen Wei Ho}
\affiliation{Department of Physics, Harvard University, Cambridge, MA 02138, USA}
\affiliation{Department of Physics, Stanford University, Stanford, CA 94305, USA}

\author{Wolfgang Ketterle}
\affiliation{Department of Physics, Massachusetts Institute of Technology, Cambridge, MA 02139, USA}
\affiliation{Research Laboratory of Electronics, Massachusetts Institute of Technology, Cambridge, MA 02139, USA}
\affiliation{MIT-Harvard Center for Ultracold Atoms, Cambridge, MA, USA}

\begin{abstract}
% We study spin dynamics of ultracold atoms in optical lattices realizing  anisotropic Heisenberg models and observe in 1D special helical spin patterns with long lifetimes. Our finding corroborates the recent prediction of phantom Bethe states, special many-body eigenstates carrying finite momenta yet no energy. We theoretically find similar stable spin helices in systems of higher spatial dimensions, spin quantum numbers, and other lattice geometries, where their long-lived nature implies non-thermalizing dynamics associated with quantum many-body scars. We propose an experimental protocol to realize these states using ultracold atoms. Lastly, we use the phantom spin helices to measure the interaction anisotropy. We find a major contribution from short-range off-site interactions in the underlying Bose-Hubbard model which have not been observed before.

Exact solutions for quantum many-body systems are rare and provide valuable insight to universal phenomena. Here we show experimentally in anisotropic Heisenberg chains that special helical spin patterns can have very long lifetimes. This finding confirms the recent prediction of phantom Bethe states, exact many-body eigenstates carrying finite momenta yet no energy. We theoretically find analogous stable spin helices in higher dimensions and in other non-integrable systems, where they imply non-thermalizing dynamics associated with quantum many-body scars. We use phantom spin helices to directly measure the interaction anisotropy which has a major contribution from short-range off-site interactions that have not been observed before. Phantom helix states open new opportunities for quantum simulations of spin physics and studies of many-body dynamics.

\end{abstract}

\maketitle

The dynamics of strongly-interacting, quantum many-body systems is an active frontier of research. It has broad implications ranging from understanding fundamental phenomena like quantum thermalization or the lack thereof \cite{Berges2004_Prethermalization, Rigol2007, Schmiedmayer2012_Prethermalization, Hadzibabic2018_PrethermalBosons, abanin2019}, to realizing new forms of matter (e.g. time crystals \cite{Choi2017_TimeCrystal,Monroe2017_TimeCrystal}) and to controlling entanglement for quantum information processing \cite{Arute2019,Flamini2019}. 
 
However, analyzing in full such systems is difficult due to their complexity. Exactly-solvable models are therefore especially important and desirable since they can directly reveal the physical mechanisms behind universal phenomena. For example, the transverse field Ising model in one spatial dimension can be solved in terms of free fermions and serves as a paradigmatic model for quantum criticality \cite{JordanWigner,Sachdev2011}.

The spin-1/2 anisotropic Heisenberg model in one dimension, which is given by
\begin{flalign}
	H &= J_{xy} \sum_{\langle i j \rangle}  [ S_i^x S_j^x + S_i^y S_j^y + \Delta S_i^z S_j^z  ],
    \label{eq:Heisenberg_eq}
\end{flalign}
is another such example.  Here, the transverse and longitudinal spin couplings (between neighboring sites $i$, $j$) are $J_{xy}$ and $J_z\,{:=}\,\Delta J_{xy}$, where $\Delta$ is the spin-exchange anisotropy. 
While the model seems simple and has been known to be solvable by the Bethe ansatz for almost a century \cite{BetheAnsatz,Franchini2017}, it gives rise to rich dynamics which are still not completely understood. Indeed, only recently was it predicted that the isotropic system ($\Delta\,{=}\,1$) yields an exotic Kardar-Parisi-Zhang superdiffusive regime of transport \cite{Znidaric2011, Ljubotina2017, KPZ_2019_theory} which has been subsequently experimentally confirmed by \cite{bloch2021_KPZ}. 
Yet another surprise has come from the recent theoretical discovery of a special set of degenerate many-body eigenstates in the model for any anisotropy --- so-called phantom Bethe states \cite{popkov2021phantom}. These are states composed of quasiparticles which carry momentum  but contribute zero energy (relative to the ferromagnetic \enquote{vacuum} state),  akin to ghost particles, hence the name phantom.
%These phantom properties carry over to 
Simple patterns of spins winding in the transverse plane --- i.e.~spin-helix states --- also share these phantom properties if their pitch $\lambda$ or wavevector $Q\,{:=}\,2\pi/\lambda$\,$=$\,$Q_p$, where $Q_p$ satisfies  the phantom condition 
\begin{align}
    \Delta = \cos(Q_p a),
    \label{eq:phantom_condition}
\end{align}
with $a$ being the lattice spacing. These states, which we call phantom helix states, are exact many-body eigenstates and do not decay. Since interactions, even in an integrable model, are expected to cause a system to locally relax to a (generalized) Gibbs ensemble \cite{Cassidy2011, Franchini2017}, such a long-lived and far-from-equilibrium state represents a surprising exception to (generalized) quantum thermalization.
 
In this work, we systematically explore the dynamics of spin-helix states using our versatile ultracold atom quantum simulator platform with tunable anisotropy \cite{Jepsen2020_SpinTransport, Jepsen2021_TransverseSpin}. Specifically, we study their decay as a function of wavevector $Q$ for different fixed anisotropies $\Delta$, and find a non-monotonic decay rate with a pronounced minimum near the expected special value $Q_p$. This is the signature of the phantom spin-helix state, confirming the predictions of \cite{popkov2021phantom}. 
We further theoretically establish generalizations of the phantom spin-helix states to  Heisenberg systems of higher dimensions, with higher spin quantum numbers, and for non-cubic lattice geometries. 
While 1D spin-1/2 models are integrable, these generalizations are not; therefore, the existence of stable far-from-equilibrium helices in such systems leads to genuinely non-thermalizing dynamics associated with quantum many-body scars \cite{Serbyn2021}. 
We propose an experimental protocol to realize them with ultracold atoms. Lastly, we demonstrate how dynamics of phantom helices can be used as an important tool for quantum simulations of spin physics.  
Using the phantom condition (\cref{eq:phantom_condition}) it is now possible to experimentally determine the anisotropy $\Delta$. We find that the anisotropy is strongly affected by nearest-neighbor (i.e.~off-site) interactions of the underlying Hubbard model, which have not been observed before for contact interactions.

\begin{figure}[t]
   \includegraphics[width=\linewidth,keepaspectratio]{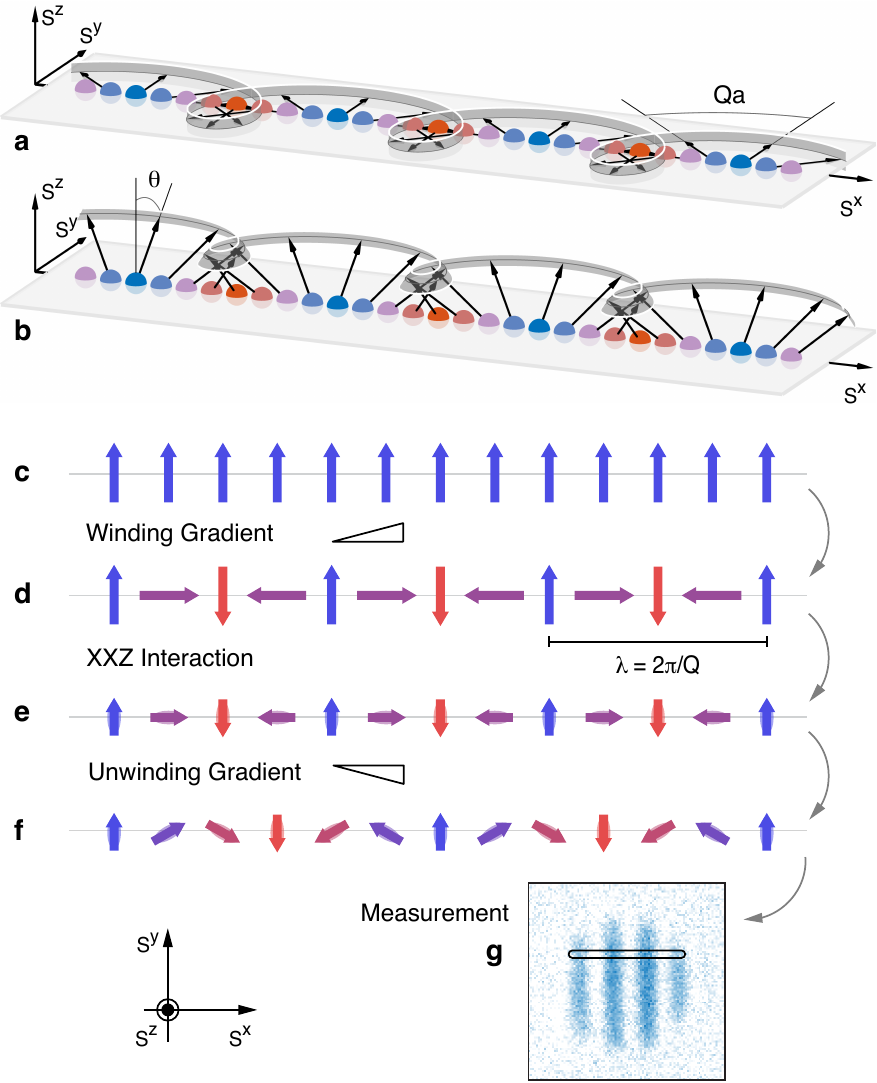}
    \caption{
        \textbf{Preparation and observation of spin-helix states}. We prepare a transverse spin helix in the $S^x$-$S^y$ plane, i.e.~$\theta\,{=}\,\pi/2$ (\textbf{a}), or with arbitrary polar angle $\theta$ (\textbf{b}). An initially spin-polarized state in the $S^y$ direction (\textbf{c}) is wound into a spin helix with variable wavevector $Q$ using a magnetic field gradient, here illustrated for $Qa\,{=}\,\pi/2$ (\textbf{d}). This state evolves under the XXZ Heisenberg Hamiltonian (\textbf{e}). After unwinding the remaining spin modulation to a resolvable wavevector (\textbf{f}), the local $S^y$-magnetization is imaged in-situ (\textbf{g}). Only the $S^x$ and $S^y$ components of the spin are shown for \textbf{c}-\textbf{f}.
        }
    \label{fig:spinHelices}
	\vspace{-5pt}
\end{figure}

\textbf{\textit{Spin-helix states \& the phantom condition.}} In this work, we study transverse spin-helix states 
\begin{align}
     \ket{\psi(Q)}=\prod_n \left[\cos(\theta/2)\ket{\uparrow}_n\,{+}\,\sin(\theta/2)e^{-i Q z_n }\ket{\downarrow}_n \right].
     \label{eq:spinspirals}
\end{align}
The polar angle $\theta$ determines the local longitudinal spin component $\langle S^z_n \rangle$ which is constant along the chain,
and $z_n$ is the position of the $n$-th spin (see \cref{fig:spinHelices}a,b). In the classical limit, {\it any} transverse spin helix is stable for {\it any} anisotropy since the torques exerted on a given spin by its neighbors cancel exactly \cite{Jepsen2021_TransverseSpin}. Therefore, the decay of a spin helix is due to quantum fluctuations. However, for wavevectors $Q_p$ fulfilling the phantom condition for $|\Delta|\,{\leq}1$ (\cref{eq:phantom_condition}), the fluctuations from the two nearest neighbors \textit{also} cancel exactly, making these special helices particularly long-lived. Intriguingly, one can show that they are in fact
many-body eigenstates of the Heisenberg model (\cref{eq:Heisenberg_eq}) for infinite systems, or for finite systems with appropriate boundary conditions \cite{popkov2021phantom}. For the finite chains with open boundaries that we prepared in our experiments, the phantom spin helix is only metastable --- defects will propagate from the ends of the chain into the bulk, resulting in a nonzero but small decay rate \cite{Jepsen2021_TransverseSpin}.
 
For the isotropic system ($\Delta\,{=}\,1$), the phantom spin helix has wavevector $Q_p\,{=}\,0$ and thus reduces to a spin-polarized product state, which is a trivial eigenstate of this model. For $\Delta\,{=}\,0$ the phantom condition gives $Q_pa\,{=}\,\pi/2$, so that angles between neighboring spins are 90$^\circ$ (assuming all spins lie in the $S^x$-$S^y$ plane, i.e.~$\theta\,{=}\,\pi/2$).
To explain intuitively how this state is metastable, consider a spin which points e.g.~in the $S^x$ direction, with quantum fluctuations in the $S^y$-$S^z$ plane. For $J_z\,{=}\,0$, there is no interaction from the $S^z$ component,
while the $S^y$ component causes no precession on the neighboring spins, which also point along the ${\pm}S^y$-direction. In the following, we will demonstrate the existence of these long-lived phantom helices for general $\Delta$, and confirm the predictions of the phantom condition \cref{eq:phantom_condition}.

\textbf{\textit{Experimental methods.}}  As in our previous work \cite{Jepsen2020_SpinTransport,Jepsen2021_TransverseSpin}, the spin model is implemented by loading ultracold lithium-7 atoms in the two lowest hyperfine states into a three-dimensional optical lattice. This system is well-described by a two-component Bose-Hubbard model. Two of the three lattice potentials are kept high at $35\,E_R$ creating a bundle of isolated 1D-chains. The lattice depth of the third axis $V_0$ is set to a value between $9\,E_R$ and $11\,E_R$, which is deep enough such that the system remains in the Mott insulating regime while still allowing for spin dynamics. Here $E_R\,{=}\,{h^2/(8ma^2)}$ is the recoil energy, $m$ the atomic mass, and $h$ the Planck constant.

Because particle motion is suppressed in the Mott insulator, the dynamics of the remaining degrees of freedom can be described using a pure spin model. By mapping the two hyperfine states onto spins $\ket{\uparrow}$ and $\ket{\downarrow}$, we can realize the spin-$1/2$ Heisenberg XXZ model (\cref{eq:Heisenberg_eq}) in which the interactions are mediated by superexchange \cite{Svistunov2003_CounterflowSF, Duan2003_ControllingSpinExchange, GarciaRipoll2003_BosonsInOpticalLattice, Altman2003_TwoComponentBosons}. The transverse and longitudinal spin couplings are given by
\begin{equation}
  \begin{aligned}
    J_z &= 4\tilde{t}^2/U_{\uparrow\downarrow}-(4\tilde{t}^2/U_{\uparrow\uparrow}+4\tilde{t}^2/U_{\downarrow\downarrow}) \\
    J_{xy} &=-4 \tilde{t}^2/U_{\uparrow\downarrow},
  \end{aligned}
  \label{eq:superexchange}
\end{equation}
where $\tilde{t}$ is the tunneling matrix element between neighboring sites and $U_{\uparrow\uparrow}$, $U_{\uparrow\downarrow}$, $U_{\downarrow\downarrow}$ are the on-site interaction energies. The spin couplings (\cref{eq:superexchange}) can be varied over two orders of magnitude by changing the lattice depth $V_0$, which scales the entire Hamiltonian. We control the anisotropy $\Delta\,{:=}\,J_z/J_{xy}$ via an applied magnetic field $B$, which tunes the interactions through Feshbach resonances \cite{interactionSpectroscopy, Jepsen2020_SpinTransport, Jepsen2021_TransverseSpin}. In our realization, the transverse coupling is antiferromagnetic ($J_{xy}\,{>}\,0$). 

The transverse spin helix is created by RF pulses to tilt the spins to a finite polar angle $\theta$, followed by magnetic field gradients to wind a helix \cite{Bloch2014_SpinHelix,  Thywissen2015_LG_effect, Jepsen2021_TransverseSpin}. Time evolution is initiated by rapidly lowering $V_0$. The dynamics following this quench are governed by the 1D XXZ model (\cref{eq:Heisenberg_eq}, \cref{fig:spinHelices}e) with a selected anisotropy $\Delta$. After a variable evolution time $t$, the dynamics are frozen by rapidly increasing $V_0$; the sample is then imaged.

Our imaging system limits the direct observation of spin modulations to a wavelength of $\lambda\,{>}\,6\,a$. In order to image spin helices at any value of $Q$, we first unwind the remaining spin modulation to a wavelength of $\lambda\,{\approx}\,10\,a$ by applying a $\pi$-pulse followed by a magnetic field gradient as in \cref{fig:spinHelices}f. After unwinding, we turn the transverse helix into a population modulation by applying a $\pi/2$-pulse. In the end, we detect the spatial distribution of spin $\ket{\uparrow}$ atoms in-situ with state-selective polarization-rotation imaging (\cref{fig:spinHelices}g). Compared to our previous work \cite{Jepsen2020_SpinTransport, Jepsen2021_TransverseSpin}, this novel unwinding step extends our observable range of wavevectors $Q$ all the way to $Qa\,{=}\,\pi$, where neighboring spins are anti-aligned. Integrating the images along a direction perpendicular to the chains yields a 1D spatial profile of sinusoidal population modulation over all spin chains (\cref{fig:spinHelices}g). As described in Refs.~\cite{Jepsen2020_SpinTransport, Jepsen2021_TransverseSpin}, we determine the normalized contrast $c(t)$ which decays during the evolution time $t$ and obtain a decay rate $\gamma$ by using a linear fit at early times.

\begin{figure}[t]
    \includegraphics[width=\linewidth,keepaspectratio]{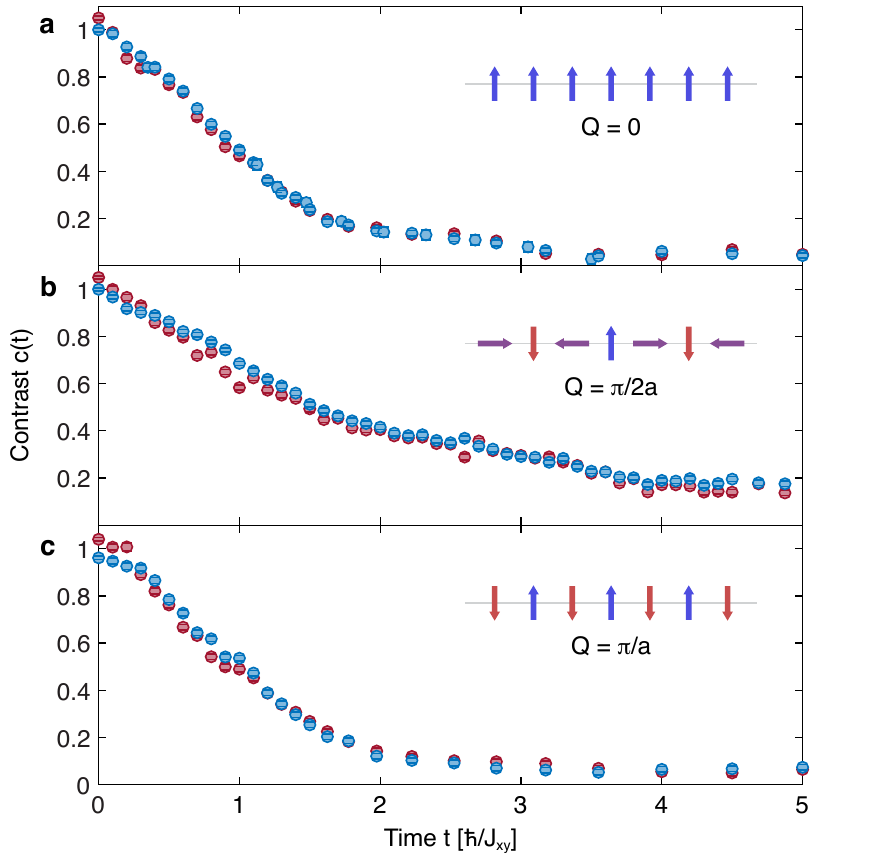}
    \caption{
        \textbf{Decay of spin-helix states}. Spin-helix contrast $c(t)$ measured for $\Delta\,{\approx}\,0$ at two different lattice depths $9\,E_R$ (red) and $11\,E_R$ (blue), for three wavevectors. \textbf{a}, $Qa\,{=}\,0$, all spins aligned. \textbf{b},  $Qa\,{=}\,\pi/2$, neighboring spins perpendicular, and an eigenstate for $\Delta\,{=}\,0$. \textbf{c}, $Qa\,{=}\,\pi$, all spins anti-aligned. Decay curves collapse for each wavevector when times are normalized in units of $\hbar/J_{xy}$. The contrast lifetime is significantly larger for $Qa\,{=}\,\pi/2$ compared to $Qa\,{=}\,0$ and $Qa\,{=}\,\pi$.
        }
    \label{fig:decayCurves}
	\vspace{-0pt}
\end{figure}

\begin{figure*}[t]
    \includegraphics[width=\linewidth,keepaspectratio]{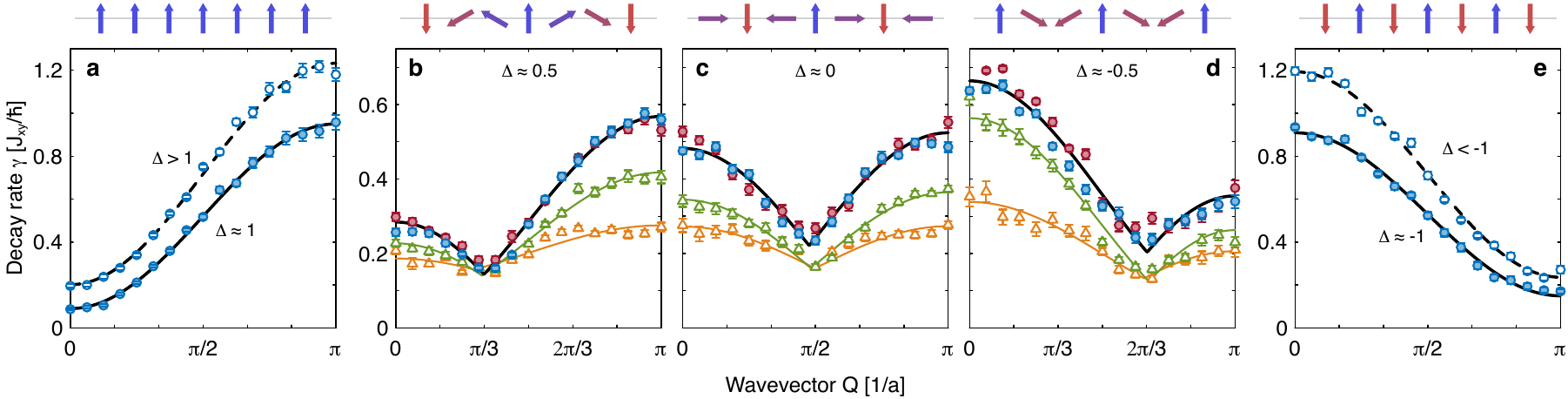}
    \caption{
        \textbf{Observation of phantom helix states}.
        Decay rate $\gamma$ as a function of wavevector $Q$, shown for fitted anisotropies $\Delta$ ranging from positive (\textbf{a}-\textbf{b}) through zero (\textbf{c}) to negative values (\textbf{d}-\textbf{e}) measured at lattice depths of $9\,E_R$ (red) and $11\,E_R$ (blue). The decay rate minimum occurs at a wavevector $Q_p$ which increases smoothly from $Q_pa\,{=}\,0$ for $\Delta\,{\approx}\,1$ (\textbf{a}) to $Q_pa\,{=}\,\pi$ for $\Delta\,{\approx}\,{-}1$ (\textbf{e}).
        The long-lived spin pattern in the $S^x$-$S^y$ plane is illustrated on top of each panel. Fits $\gamma(Q)\,{=}\,\gamma_1|\Delta\,{-}\,\cos(Qa)|\,{+}\,\gamma_0$ (lines) are used to find $Q_p$ and the anisotropies $\Delta$ shown in \cref{fig:anisotropyVsField}. In addition to the purely transverse helices (polar angle $\theta\,{=}\,\pi/2$, circles), panels (\textbf{b}-\textbf{d}) also show the decay of spin helices with polar angles $\theta\,{=}\,5\pi/12$ (green triangles) and $\theta\,{=}\,2\pi/3$ (orange triangles).  The curves above were measured at applied magnetic fields of $B\,{=}\,847.887\,\text{G}$ ($\Delta\,{>}\,1$) and $847.286\,\text{G}$ ($\Delta\,{\approx}\,1$) (\textbf{a}), $845.760\,\text{G}$ (\textbf{b}), $842.905\,\text{G}$ (\textbf{c}), $839.376\,\text{G}$ (\textbf{d}), $833.004\,\text{G}$ ($\Delta\,{\approx}\,{-}1$) and $827.287\,\text{G}$ ($\Delta\,{<}\,{-}1$) (\textbf{e}).
        }
    \label{fig:gammaVsQ}
	\vspace{-0pt}
\end{figure*}

\textbf{\textit{Experimental observations of phantom helices.}}  \cref{fig:decayCurves} illustrates the contrast decay $c(t)$ for spin helices with different wavevectors $Q$ at $\Delta\,{\approx}\,0$. We see that the decay for $Qa\,{=}\,\pi/2$ is noticeably slower than that of $Qa\,{=}\,0$ or $\pi$. This non-monotonic behavior is the signature of phantom spin helices. This feature is emphasized in \cref{fig:gammaVsQ} by comparing the contrast decay rates $\gamma$ at various values of $Q$. According to the phantom condition \cref{eq:phantom_condition}, the wavevector $Q_p$ with the smallest decay rate changes as a function of $\Delta$. Since the superexchange interactions in \cref{eq:superexchange} depend on the scattering lengths, we can tune $\Delta$ smoothly by varying the magnetic field. Indeed, in \cref{fig:gammaVsQ} we see $Q_p$ varies accordingly as predicted.

\begin{figure*}[t]
    \includegraphics[width=\linewidth,keepaspectratio]{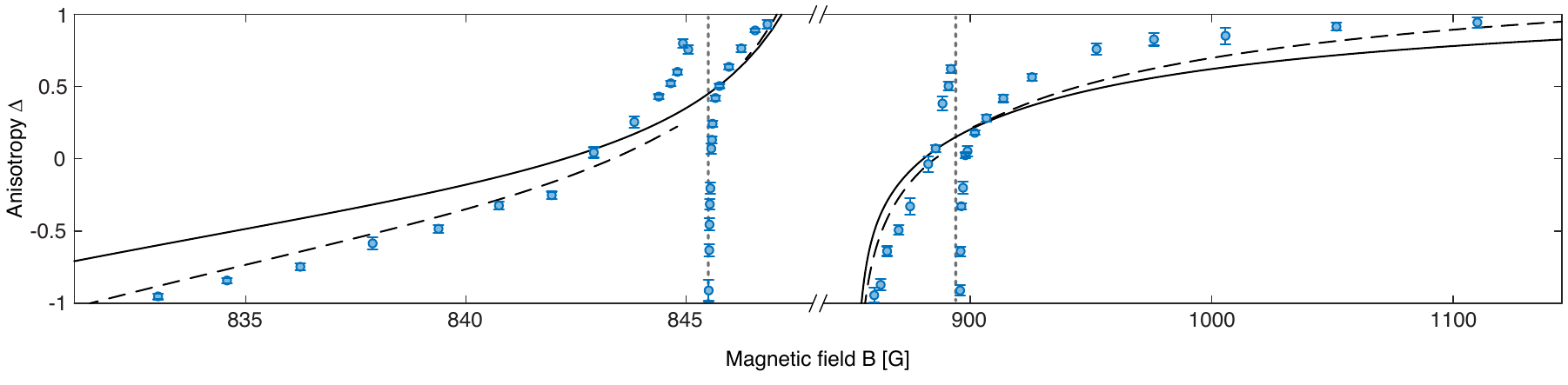}
    \caption{
        \textbf{Tuning the anisotropy with magnetic fields}.  Measured anisotropies $\Delta$ are compared to the standard model for superexchange given by \cref{eq:superexchange} (solid line) using previously measured scattering lengths \cite{Kokkelmans2020_interactionSpectroscopy}. The black dashed line includes the bond-charge correction to tunneling \cite{Lhmann2012} and a small adjustment of the background scattering length of $a_{\uparrow\downarrow}$, which was not tightly constrained by previous measurements. Major deviations near the two Feshbach resonances at 845.506 and 893.984 G (vertical dotted lines) are evidence for off-site interactions.}
    \label{fig:setup}
	\vspace{-5pt}
	\label{fig:anisotropyVsField}
\end{figure*}

The fit function $\gamma(Q)\,{=}\,\gamma_1|\Delta\,{-}\,\cos(Qa)|\,{+}\,\gamma_0$ was derived from a short-time expansion of the spin-helix contrast $c(t)$~\cite{Jepsen2021_TransverseSpin}. Here $\Delta$, $\gamma_1$ and $\gamma_0$ are treated as free fit parameters, where $\gamma_0$ represents a background decay rate accounting for effects such as finite chain length, holes in the spin chains, and inhomogeneous dephasing \cite{Jepsen2021_TransverseSpin}. The predicted $\Delta$ based on \cref{eq:superexchange} and previously determined scattering lengths  \cite{Kokkelmans2020_interactionSpectroscopy} agrees qualitatively with the $\Delta$ we fit (see \cref{fig:anisotropyVsField}). We also observe phantom helix states for various polar angles in \cref{fig:gammaVsQ}b-d. This confirms the prediction \cite{popkov2021phantom} that there is a whole family of phantom helix states for a given value of $\Delta$. We find that a larger absolute value of the total $S^z$ magnetization (i.e. $\theta$ close to $0$ or $\pi$) leads to an overall slower decay.

For anisotropies $|\Delta|\,{>}\,1$ (\cref{fig:gammaVsQ}a,e; open symbols) there is no longer a stable spin-helix eigenstate \cite{popkov2021phantom}. We see instead that the minimum decay rate is always at $Q\,{=}\,0$ for $\Delta\,{\geq}\,1$ and $Qa\,{=}\,\pi$ for $\Delta\,{\leq}\,{-}1$. Comparing decay rates across $\Delta$ in this range, we find a $Q$-independent increase relative to the $|\Delta|\,{=}\,1$ case which is monotonous in $|\Delta|\,{-}\,1$.

Our improved imaging protocol allows us to access new parameter regimes beyond previous work. For the isotropic system ($\Delta\,{=}\,1$) we had observed diffusive spin transport, characterized by $\gamma(Q)\,{\approx}\,\gamma_1 Q^2 a^2/2\,{+}\,\gamma_0$ \cite{Bloch2014_SpinHelix, Jepsen2020_SpinTransport, Jepsen2021_TransverseSpin}. However, we now see this quadratic behavior break down for large $Q$ when the wavelength $\lambda$ becomes comparable to the lattice spacing $a$. The fastest decay occurs for the N\'eel state ($Qa\,{=}\,\pi$), where neighbouring spins are anti-aligned. This directly demonstrates that this \textit{classical} antiferromagnetic state is \textit{not} the ground state of the \textit{quantum} antiferromagnetic Heisenberg Hamiltonian. Nevertheless, the N\'eel state is an exact highly-excited eigenstate for $\Delta\,{=}\,{-}1$ (\cref{fig:gammaVsQ}e).

\textbf{\textit{Extension to higher dimensions.}} The phantom spin-helix states were originally discovered in 1D as a coherent superposition of phantom Bethe states, special degenerate solutions to the Bethe ansatz equations \cite{popkov2021phantom}. This raises the question if they are a phenomenon tied exclusively to integrability. We find that they are not: we can show that stationary phantom helix states exist for any anisotropy, for the  anisotropic Heisenberg model defined on hypercubic lattices in {\it arbitrary} spatial dimensions and {\it arbitrary} spin quantum numbers. Specifically, we consider the generalization of the spin-helix states (see Supplementary Materials) in $d$ dimensions specified by a wavevector $\mathbf{Q}\,{=}\,(Q_1,\cdots,Q_d)$, and only require that $\mathbf{Q}\,{=}\,\mathbf{Q}_p\,{:=}\,Q_p \mathbf{x}$, where $\mathbf{x}\,{\in}\,\{-1,1\}^d$ is a binary vector of $-1$s and $1$s, and $Q_p$ ($0\,{\leq}\,Q_p a\,{\leq}\,\pi$) satisfies the phantom condition (\cref{eq:phantom_condition}). For example, for the Heisenberg model with $\Delta\,{=}\,1/2$ on a 2D square lattice so that $Q_p a\,{=}\,\pi/3$, there are four such wavevectors:~$\mathbf{Q}_p a\,{=}\, (\pi/3,\pi/3)$, $(\pi/3,-\pi/3)$, $(-\pi/3,\pi/3)$, $(-\pi/3,-\pi/3)$; all such states are phantom helices (see Fig.~\ref{fig:square}). In fact, non-trivial phantom helix states also exist for non-hypercubic lattices, e.g.~triangular and kagome lattices (see Fig.~\ref{fig:2D_generalized_lattices}), but only for the special value of the anisotropy  $\Delta\,{=}\,{-}1/2$. We prove these statements % \sout{and provide more details}
in the Supplementary Material.

Our experimental protocol can be extended to two or three dimensions to directly observe quantum scarred dynamics associated with these special helices. An experimental complication is that near Feshbach resonances, any  magnetic field gradient is projected along the direction of the strong bias field. Therefore, to wind a helix with arbitrary $\mathbf{Q}\,{=}\,(Q_x,Q_y,Q_z)$ vectors, one must load atoms into a deep 3D lattice, ramp to low fields $B$ where appropriate gradients in any direction can be created easily, wind the transverse spin helix with arbitrary $\mathbf{Q}$, and then return to the high fields near Feshbach resonances. Because of the small scattering lengths of lithium-7 at low field, a very deep optical lattice is required to stay in the Mott insulator regime.

\textbf{\textit{Measurement of Anisotropy.}} Besides being of fundamental interest, the phantom helix states also have practical applications. We can use the sensitivity of the phantom helix states to measure the spin-exchange anisotropy $\Delta\,{:=}\,J_z/J_{xy}$ precisely as a function of applied magnetic field $B$ (shown in \cref{fig:anisotropyVsField}). Until now, there has been no protocol to directly measure $\Delta$; it could only be derived \cite{Jepsen2020_SpinTransport, Jepsen2021_TransverseSpin} from measured scattering lengths $a_{\uparrow\uparrow}$, $a_{\uparrow\downarrow}$, $a_{\downarrow\downarrow}$ \cite{interactionSpectroscopy, Kokkelmans2020_interactionSpectroscopy} using \cref{eq:superexchange}. \cref{fig:anisotropyVsField} compares our measured $\Delta$ to the predictions based on scattering lengths. They agree quite well far from the Feshbach resonances. We will first discuss the accuracy of our determination of $\Delta$ and possible systematic errors, then comment on the new physics near Feshbach resonances.

Repeated measurements of $\Delta$ have a reproducibility of better than 0.1, clearly visible by the small random scatter of the data points in \cref{fig:anisotropyVsField}. In addition, there are several possible systematic effects. (1) The lineshapes of contrast decay are non-exponential and differ between short and long decay times (visible in \cref{fig:decayCurves}), partly due to a beat note between the precession of atoms in the spin chains and isolated atoms in the outer part of the cloud \cite{Jepsen2021_TransverseSpin}. The resulting oscillations average out for long decay times, but not for shorter decay times. By using different fit functions, we estimate that $|\Delta|$ in \cref{fig:anisotropyVsField} could be overestimated by at most ${\sim}\,0.1$. 

(2) Our fit function $\gamma(Q)\,{=}\,\gamma_1|\Delta\,{-}\,\cos(Qa)|\,{+}\,\gamma_0$ was derived from a short-time quadratic expansion of the decay of contrast \cite{Jepsen2021_TransverseSpin}. However, experimentally, we can only observe the contrast at intermediate times where the decay is more linear (as already discussed in previous work \cite{Bloch2014_SpinHelix, Jepsen2020_SpinTransport, Jepsen2021_TransverseSpin}). Numerical simulations reveal that different methods of extracting $\gamma$ can lead to a systematic overestimation of $|\Delta|$ by up to ${\sim}\,0.15$. The effect is maximum around $|\Delta|\,{\approx}\,0.5$, and falls off for $|\Delta|$ closer to 0 or 1.

(3) A systematic shift of $|\Delta|$ would also occur if there were another $Q$-dependent decay mechanism due to background decay (e.g. holes) or decay propagating from the ends of the chain.  Several of these issues can be addressed experimentally in the future by using a quantum gas microscope and observing dynamics in single isolated spin chains.

Some of the discrepancies in \cref{fig:anisotropyVsField} can be explained by corrections to the underlying standard Hubbard model \cite{Lhmann2012, Dutta2015, Jepsen2020_SpinTransport} which include density-dependent tunneling \cite{Juergensen2014_DensityInducedTunneling}, higher-band corrections to $U$ \cite{campbell2006_image_clock_shift, Will2010_MultiBodyInteractions}, and off-site contact interactions \cite{Jepsen2020_SpinTransport}. However, only contributions from off-site interactions get large near a Feshbach resonance. Therefore, the phantom helix states reveal that spin-spin interactions near a Feshbach resonance are dominated by off-site interactions which have never been observed for contact interactions\footnote{Off-site interactions were observed for long-range dipolar interactions \cite{Baier2016}.}. Off-site interactions originate from the small overlap of a Wannier function on one site with those of its neighbours \cite{Hirsch1996_offsiteInteractions} and add a correction term to $J_z$ in \cref{eq:superexchange} of $ 2(V_{\uparrow\uparrow}\,{+}\,V_{\downarrow\downarrow}\,{-}\,2V_{\uparrow\downarrow})$ \cite{Jepsen2020_SpinTransport} where $V_{\uparrow\uparrow}$, $V_{\uparrow\downarrow}$, $V_{\downarrow\downarrow}$ are the off-site interaction energies. In a forthcoming publication, we will show that the current model for off-site interactions \cite{Hirsch1996_offsiteInteractions, Lhmann2012, Dutta2015} is insufficient near a Feshbach resonance. 

\textbf{\textit{Discussion and outlook.}} Previous studies of the Heisenberg Hamiltonian have focused on the ground state \cite{ Greiner2016_FermiSpinCorrelations, Cheuk2016_SpinCorrelations, Greiner2017_FermiAntiferromagnet}, low-lying elementary excitations including magnons \cite{Bloch2013_SingleSpin, Bloch2013_BoundMagnons} and Bethe strings \cite{Wang2018_BetheStrings, Bera2020_BetheStringsDispersion}, or on \textit{unstable} dynamics far away from equilibrium \cite{Bloch2014_SpinHelix, Zwierlein2019_SpinTransport, Jepsen2020_SpinTransport, bloch2021_KPZ, Jepsen2021_TransverseSpin}. This work captures a new class of excitations for the Heisenberg model: the phantom spin helices. These are  highly-excited yet long-lived \textit{metastable} states. Their stability is not a result of symmetry, but rather due to a delicate cancellation of interactions.
 
We have theoretically explored spin-helix states in systems in higher dimensions, for different spin quantum numbers, and in various lattice geometries, which are not integrable.  We find that for the special initial condition of a phantom helix state, the system does not relax to thermal equilibrium as one would na\"{i}vely expect, despite the presence of strong interactions. Such dynamics in non-integrable, many-body systems constitute examples of \enquote{weak ergodicity-breaking}, or what are now known as \enquote{quantum many-body scars} \cite{Serbyn2021}. We note that while various toy models hosting exact quantum many-body scars have already been discussed in the literature, such models are primarily theoretical constructs which are difficult to realize experimentally \cite{shiraishi2017, mark2020, chattopadhyay2020, moudgalya2020}. In contrast, we demonstrate that one of the simplest examples of a many-body system (the XXZ Heisenberg model) can support quantum many-body scars, a fact which has been overlooked thus far. The simplicity implies that probing such scarred dynamics experimentally is relatively straightforward, and we propose a way to observe them with an extension of our current experimental setup to higher dimensions.

We expect phantom helix states to have applications for quantum simulations of spin physics. We have demonstrated the potential of the phantom helix states as a sensitive tool to directly measure the anisotropy $\Delta$. They have revealed that even short-range interactions can lead to strong off-site interactions in spin models.  This can now be used to realize extended Hubbard models \cite{Hirsch1996_offsiteInteractions, Dutta2015} including the quantum lattice gas (or $t$-$V$) model \cite{1_RevModPhys.83.1405} which supports a supersolid phase \cite{matsuda1970}.
In the future, these long-lived helix states could be an intermediate step in preparing other many-body quantum states or be used for robust quantum sensing \cite{Dooley2021}.  An intriguing question is what will ultimately limit the stability of these states if periodic boundary conditions are realized with ring-shaped atom arrays \cite{rydberg_XXZ_scholl2021}.  Such studies are likely to provide new insight into the rich dynamics of Heisenberg spin models.

\textbf{Acknowledgements.} We thank J.~Rodriguez-Nieva for helpful discussions, and J.  Amato-Grill for important advice. We thank Jinggang Xiang for experimental assistance, Julius de Hond for comments on the manuscript, and M. Zwierlein for sharing equipment. \textbf{Funding:} We acknowledge support from the NSF through the Center for Ultracold Atoms and Grant No. 1506369, the Vannevar-Bush Faculty Fellowship, and DARPA (grant number W911NF2010090). W.W.H.~is supported in part by the Stanford Institute of Theoretical Physics. \textbf{Author contributions:} P.~N.~J., Y.~K.~L., H.~L., W.~W.~H. and W.~K. conceived the experiment. P.~N.~J., Y.~K.~L., H.~L., I.~D., and Y.~M. performed the experiment. P.~N.~J., Y.~K.~L., and H.~L. analyzed the data. W.~W.~H. generalized the findings to higher dimensions. All authors discussed the results and contributed to the writing of the manuscript. \textbf{Competing
interests:} The authors declare no competing financial interests. \textbf{Data and materials availability:} The data that support the findings of this study are available from the corresponding author upon reasonable request.

\newpage

\clearpage

\setcounter{figure}{0}
\makeatletter 
\renewcommand{\thefigure}{S\@arabic\c@figure}
\makeatother

\onecolumngrid
\section*{Supplementary material}

\section{Phantom spin helices  in higher-dimensions, arbitrary spin quantum numbers, and non-hypercubic geometries}
 \label{Appendix:phantom_higher_dim}
In this section we extend the phenomenology of stable phantom helices to Heisenberg models in higher dimensions, arbitrary spin quantum numbers, and non-hypercubic geometries. Specifically, we prove that for any given anisotropy in the easy-plane $|\Delta|\,{\leq}\,1$, there exist phantom helices which are {\it exact} many-body eigenstates of the model, provided (i) the phantom condition (\cref{eq:phantom_condition} of the main text) holds; and (ii) appropriate boundary conditions are taken.

\subsection{Model and spin-helix states}

We consider the quantum Heisenberg XXZ model for any spatial dimensionality $d$, spin-$S$, and lattice geometry. The Hamiltonian is given by a sum over pairwise nearest-neighbor interactions (we set $J_{xy}\,{=}\,1$ for simplicity):
\begin{align}
    H(\Delta): = \sum_{\langle ij \rangle} (S^x_i S^x_j + S^y_i S^y_j) + \Delta S^z_i S^z_j,
\end{align}
where $S_i^\alpha$ ($\alpha\,{=}\,x,y,z$) are spin-$S$ operators. 

The generalization of a spin-helix state from the spin-1/2 case (\cref{eq:spinspirals} of the main text) is given by
\begin{align}
    |\psi(\mathbf{Q})\rangle = \prod_i  \left[ e^{-i \mathbf{Q} \cdot \mathbf{r}_i S^z_i } e^{-i \theta S^y_i} |S \rangle_i \right].
    \label{eq:spinspiral_generalization}
\end{align}
Here $|S\rangle_i$ is the local maximal spin state satisfying $(\mathbf{S}_i\,{\cdot}\,\mathbf{S}_i) |S\rangle_i\,{=}\,S(S+1) |S\rangle_i$
and $S^z_i | S\rangle_i\,{=}\,S | S \rangle_i$. 
$\mathbf{Q}\,{=}\,(Q_1,\cdots,Q_d)$ is a $d$-dimensional wavevector parameterizing the winding rate and direction of the spiral, and $\mathbf{r}_i$ is the coordinate of the spin at site $i$. Therefore, Eq.~\eqref{eq:spinspiral_generalization} locally describes a state created by rotation by angle $\theta$ from the $z$-polarized state around the $S^y$-axis, before a winding in the $S^x$-$S^y$ by a site-dependent angle $\mathbf{Q}\,{\cdot}\,\mathbf{r}_i$.

\subsection{Statement of results}
For the model and state above, we have the following statements (we set the lattice spacing to $a\,{=}\,1$ for brevity):\\

\noindent {\bf Theorem 1 (phantom spin helices for the hypercubic lattice)}. {\it Consider a $d$-dimensional hypercubic lattice of volume $L_1\,{\times}\,L_2\,{\times}\,{...}\,{\times}\,L_d$ and the anisotropic spin-$S$ Heisenberg model with anisotropy $\Delta$ defined on it. Let $|\Delta|\,{\leq}\,1$, which we parameterize as $\Delta\,{=}\,\cos(Q_p)$ (the \enquote{phantom condition}) for $Q_p\,{\in}\,[0,\pi)$. Define the wavevector $\mathbf{Q}_p\,{:=}\,Q_p \mathbf{x}$ where $\mathbf{x}$  is a binary vector $\mathbf{x}\,{=}\,(x_1,\cdots,x_d)\,{\in}\,\{-1,1\}^d$, and suppose that the linear dimensions $L_i$ are such that $L_i\,{=}\,2\pi m_i/Q_p$ for some $m_i\,{\in}\,\mathbb{N}$. Then, the spin helix with wavevector $\mathbf{Q}_p$ and any polar angle $\theta$ is an exact eigenstate of the model with energy $E\,{=}\,S^2 \cos(Q_p)\,{\times}\,{\#}\text{Links}$.}
\\\\
\noindent {\bf Theorem 2 (phantom spin helices for the 2D triangular and kagome lattices)}. {\it Consider a $2$-dimensional regular triangular or kagome lattice, and the anisotropic spin-$S$ Heisenberg model with anisotropy $\Delta\,{=}\,{-}1/2$ defined on it. Define $Q_p\,{=}\,2\pi/3$ so that $\cos(2\pi/3)\,{=}\,{-}1/2\,{=}\,\Delta$. Then the spin helix with wavevector $\mathbf{Q}_p\,{=}\,(2 Q_p,0)$ or $(4 Q_p, 0)$ and any polar angle $\theta$, is an exact eigenstate of the model with energy $E\,{=}\,S^2 \cos(Q_p)\,{\times}\,{\#}\text{Links}$, provided the lattice dimensions are such that it is commensurate with the spiral pattern.}
\\

\begin{figure*}[t]
    \includegraphics[width=0.5\linewidth]{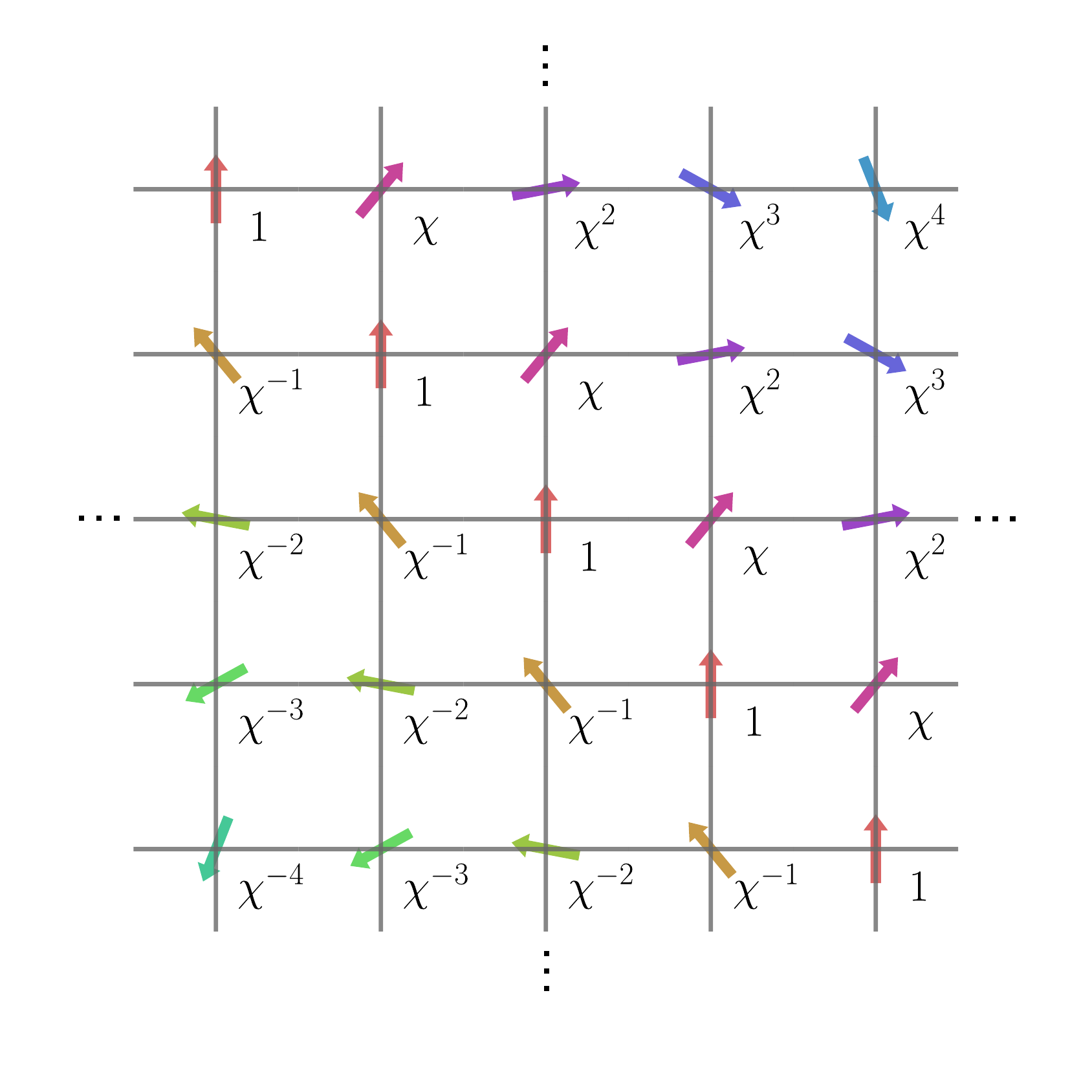}
    \caption{
        \textbf{Phantom spin helix for the square lattice.}
        Each vertex carries a phase, which denotes the state's local expectation value $\langle S^+_i\rangle\,{=}\,\langle S^x_i\rangle\,{+}\,i \langle S^y_i \rangle$. Here,  $\chi\,{:=}\,e^{-iQ_p}$. % , and each spin has a pair of neighbors with relative angle $\pm Qp$ in $S^x$-$S^y$ plane in each direction. 
        We have %set $a=1$ for simplicity and
        assumed the spiral lies fully in the transverse plane (polar angle $\theta\,{=}\,\pi/2$), but the latter condition can be straightforwardly lifted to allow for arbitrary polar angles. Thus, the depicted helix has the particular wavevector $\mathbf{Q}\,{=}\,(Q_p, Q_p)$, where $\Delta\,{=}\,\cos(Q_p)$ ($0\,{\leq}\,Q_p\,{\leq}\,\pi$). Other valid phantom spin-spirals for this geometry are those with wavevectors satisfying $\mathbf{Q}_p\,{=}\,Q_p \mathbf{x}$, where $\mathbf{x}\,{\in}\,\{-1,1\}^2$.
        }
	\vspace{-0pt}
	\label{fig:square}
\end{figure*}

\begin{figure*}[t]

    \includegraphics[width=0.5\linewidth]{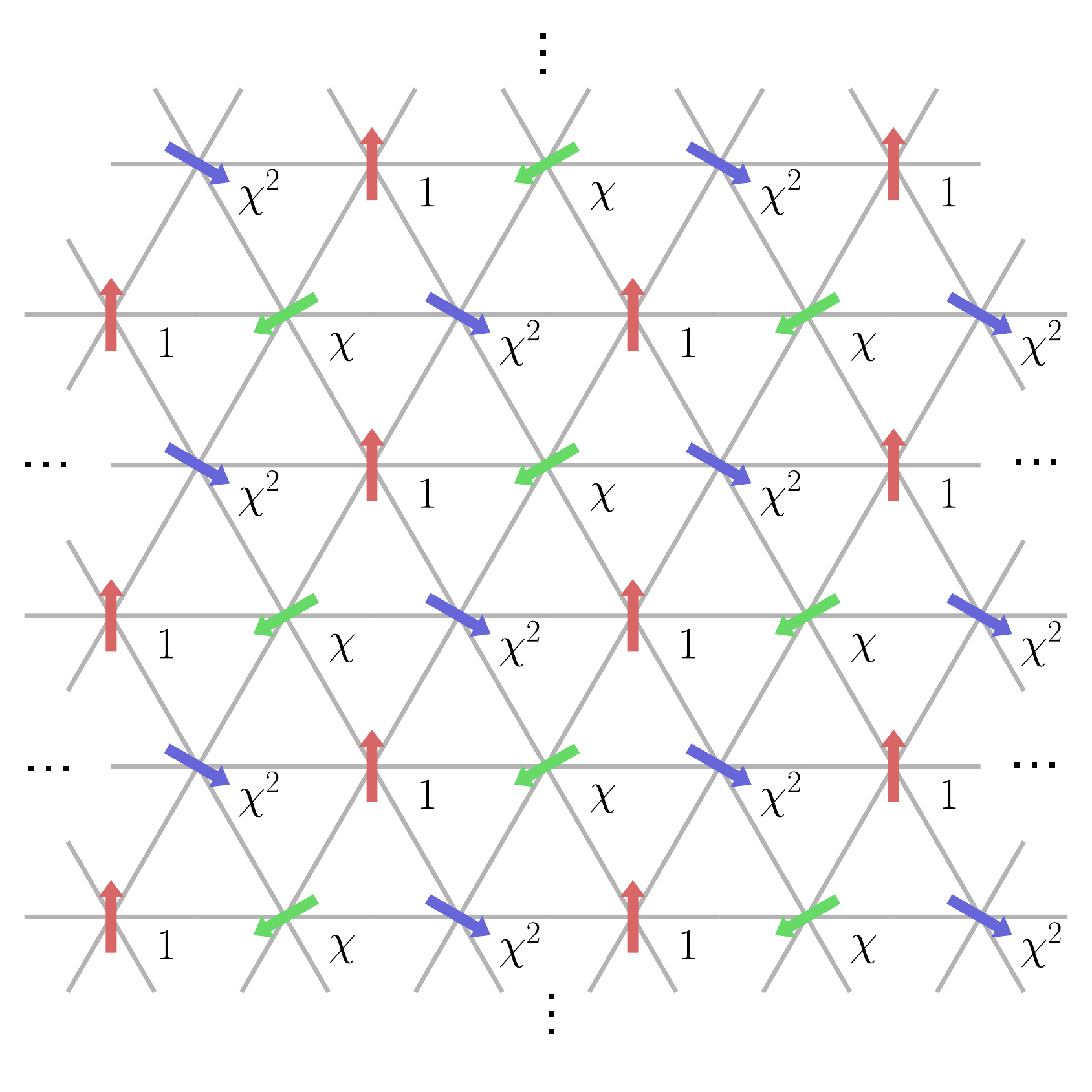}
    
    \includegraphics[width=0.55\linewidth]{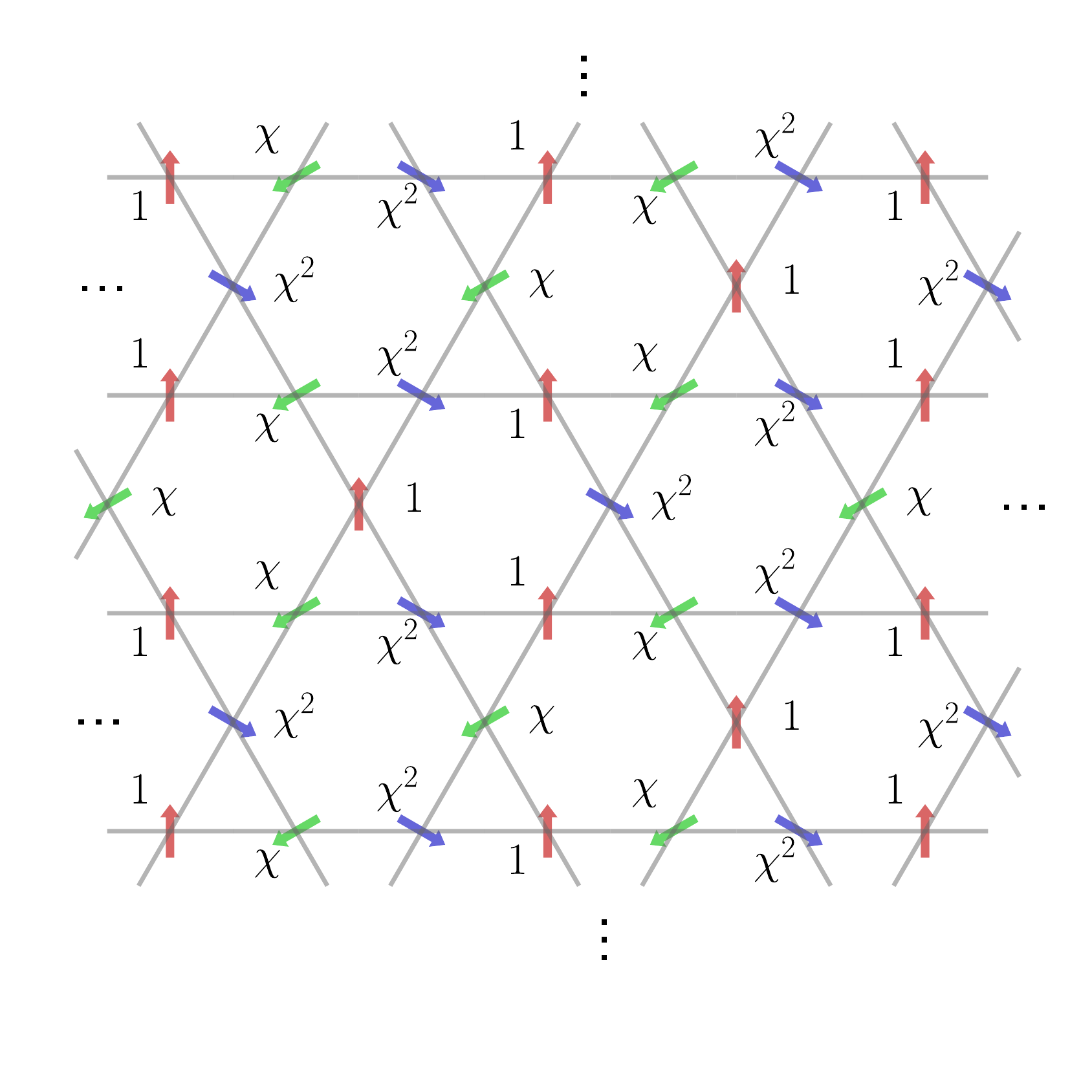}
    \caption{
        \textbf{Phantom spin helices for triangular and kagome lattices.} As in the case for the square lattice, we label each vertex with a phase denoting the local expectation value $\langle S^+_i\rangle$, defining  $\chi\,{:=}\,e^{-iQ_p}$. However, in this case, only $Q_p\,{=}\,2\pi/3$ and $\Delta\,{=}\,{-}1/2$ defines a valid phantom spin helix.  Thus, the spin can only point in one of three directions in the $S^x$-$S^y$ plane (denoted by three different colors), with relative angle $2\pi/3$ between them. Note that the collinear neighbors of each red spin are blue and green spins, and the interactions of the blue and green spins with the red spin cancel for each line. In this way, one can understand the 2D phantom helix states as arising from simple \enquote{stacking} together of phantom helices of 1D chains.
    }
	\vspace{-0pt}
	\label{fig:2D_generalized_lattices}
\end{figure*}

\noindent {\it Remarks.} 
Figure \ref{fig:square} shows an illustration of the phantom helix for a 2D square lattice, while figure \ref{fig:2D_generalized_lattices} shows the phantom helix for the 2D triangular and kagome lattices. Note the phantom helix state for the triangular lattice is  known as the 120$^{\text{o}}$ N\'{e}el ordered state in condensed matter physics, which is the ground state of the classical antiferromagnetic Heisenberg model. We show here that it is an exact eigenstate for the quantum Heisenberg model for $\Delta\,{=}\,{-}1/2$. For the kagome lattice, the phantom helix state it is also known as the $\sqrt{3}\,{\times}\sqrt{3}$ state.
\\

\noindent {\it Proof}. 
Define the local term of the Hamiltonian
\begin{align}
    h_{ij} = (S^x_i S^x_j + S^y_i S^y_j) + \cos(Q_p) S^z_i S^z_j
\end{align}
so that $H(\cos(Q_p))\,{=}\,\sum_{\langle ij\rangle} h_{ij}$. Consider a local part of the spin-helix state with wavevector $\mathbf{Q}_p$, specifically on a pair of neighboring spins $i,j$:
\begin{align}
    |\psi\rangle_{ij} & = e^{-i \Phi_i S^z_i} e^{-i \theta S^y_i} |S\rangle_i \otimes e^{-i \Phi_j S^z_j} e^{-i \theta S^y_j} |S\rangle_j \\
    & = e^{-i \Phi_i (S^z_i+S^z_j)} \left( e^{-i \theta S^y_i} |S\rangle_i \otimes e^{-i \delta \Phi_{ji} S^z_j} e^{-i \theta S^y_j} |S\rangle_j \right)
\end{align}
where $\Phi_i\,{=}\,\mathbf{Q}\,{\cdot}\,\mathbf{r}_i$ and $\delta \Phi_{ji}\,{=}\,(\Phi_j\,{-}\,\Phi_i)$. A simple but key property is that for the set-ups described in either Theorem, $e^{-i \delta \Phi_{ji}}\,{=}\,e^{-i Q_p}$ or $e^{i Q_p}$ for {\it any} nearest-neighbor pair of sites. (Clearly this is true for any hypercubic lattice. For the triangle lattice with primitive vectors $\mathbf{b}_1\,{=}\,(1,0)$, $\mathbf{b}_2\,{=}\,(1/2,\sqrt{3}/2)$ we have $e^{\pm i \frac{ 4\pi}{3}}\,{=}\,e^{\mp i Q_p}$. For the kagome lattice the primitive vectors are $\mathbf{b}_1\,{=}\,(2,0)$ and $\mathbf{b}_2\,{=}\,(1,\sqrt{3})$ while the lattice vectors within each unit cell are $\mathbf{a}_1\,{=}\,\frac{1}{2} \mathbf{b}_1$, $\mathbf{a}_2\,{=}\,\frac{1}{2} \mathbf{b}_2$ and so the result is the same as the triangular lattice).

With this in mind, we evaluate the action of $h_{ij}$ on the state (dropping indices $i,j$ for brevity)
\begin{align}
    &e^{-i \Phi_i(S^z \otimes \mathbb{I} + \mathbb{I}\otimes S^z)} (S^x \otimes S^x + S^y \otimes S^y + \cos(Q_p) S^z \otimes S^z) \left[ e^{-i \theta S^y} |S\rangle \otimes e^{\mp i Q_p S^z} e^{-i \theta S^y} |S\rangle \right] \nonumber \\
    =  &  e^{-i \Phi_i(S^z \otimes \mathbb{I} + \mathbb{I}\otimes S^z)} (\mathbb{I} \otimes e^{\mp i Q_p S^z}) \left(S^x \otimes (S^x \cos(Q_p) \mp S^y \sin(Q_p)) + S^y\otimes (S^y \cos(Q_p) \pm S^x \sin(Q_p)) + \cos(Q_p) S^z \otimes S^z \right) \nonumber \\
    & \times \left[ e^{-i \theta S^y} |S\rangle \otimes   e^{-i \theta S^y} |S\rangle \right] \nonumber \\
    = & e^{-i \Phi_i(S^z \otimes \mathbb{I} + \mathbb{I}\otimes S^z)} (\mathbb{I} \otimes e^{\mp i Q_p S^z}) \left( \cos(Q_p) (S^x\otimes S^x + S^y \otimes S^y + S^z \otimes S^z ) \mp \sin(Q_p) ( S^x \otimes S^y - S^y \otimes S^x)   \right) \nonumber \\
    & \times \left[ e^{-i \theta S^y} |S\rangle \otimes   e^{-i \theta S^y} | S\rangle \right].
    \end{align}
The term proportional to $\cos(Q_p)$ is $\mathbf{S}_i\,{\cdot}\,\mathbf{S}_j$, and the state on the RHS of the last line of the above equation is a uniformly polarized state, so it evaluates to the original state up to a multiplicative factor
\begin{align}
 S^2 \cos(Q_p) |\psi\rangle_{ij}.
\end{align}
Now we just have to evaluate the term proportional to $\sin(Q_p)$. Ignoring the factor $e^{-i \Phi_i(S^z \otimes \mathbb{I} + \mathbb{I}\otimes S^z)} (\mathbb{I}\,{\otimes}\,e^{\mp i Q_p S^z})$ we have
\begin{align}
     & \mp \sin(Q_p) ( S^x \otimes S^y - S^y \otimes S^x)     \left[ e^{-i \theta S^y} |S\rangle \otimes   e^{-i \theta S^y} |S\rangle \right] \nonumber \\
     = & \mp \sin(Q_p) (e^{-i \theta S^y} \otimes e^{-i\theta S^y} ) \left( (S^x \cos(\theta) + S^z \sin(\theta)) \otimes S^y - S^y \otimes (S^x \cos(\theta) + S^z \sin(\theta) ) \right)   |S \rangle \otimes |S\rangle   \nonumber \\
     = &  \mp \sin(Q_p) (e^{-i \theta S^y} \otimes e^{-i\theta S^y} ) \left(  \cos(\theta) (S^x \otimes S^y - S^y \otimes S^x) + \sin(\theta)S (\mathbb{I} \otimes S^y - S^y \otimes \mathbb{I} )\right) |S \rangle \otimes |S\rangle.
     \label{eq:term_propto_sinQ}
\end{align}
Now we make use of the fact that
\begin{align}
    S^y |S\rangle = i S^x |S \rangle
\end{align}
(this follows from the definition of $|S\rangle$ as the highest-weight state, $S^x\,{=}\,\frac{1}{2}(S^+\,{+}\,S^-)$, and $S^y\,{=}\,\frac{1}{2i}(S^+\,{-}\,S^-)$).

Therefore
\begin{align}
    (S^x \otimes S^y - S^y \otimes S^x) |S \rangle \otimes |S\rangle = 0
\end{align}
and Eq.~\eqref{eq:term_propto_sinQ} becomes
\begin{align}
    \mp \sin(Q_p) (e^{-i \theta S^y} \otimes e^{-i\theta S^y} ) i \sin(\theta) S (\mathbb{I} \otimes S^x - S^x \otimes \mathbb{I} ) |S\rangle \otimes | S\rangle.
\end{align}
We add a trivial term
\begin{align}
     \mp \sin(Q_p) (e^{-i \theta S^y} \otimes e^{-i\theta S^y} ) i \cos(\theta) S  (\mathbb{I} \otimes S^z - S^z \otimes \mathbb{I} ) |S\rangle \otimes |S\rangle
\end{align}
 to it, so that Eq.~\eqref{eq:term_propto_sinQ} is equal to
\begin{align}
    \pm i S \sin(Q_p) ( \mathbb{I} \otimes S^z - S^z \otimes \mathbb{I}) (e^{-i \theta S^y} \otimes e^{-i\theta S^y} )|S\rangle \otimes |  S\rangle.
\end{align}
Reinstating the factor $e^{-i \Phi_i(S^z \otimes \mathbb{I} + \mathbb{I}\otimes S^z)} (\mathbb{I}\,{\otimes}\,e^{\mp i Q_p S^z})$ which commutes with $( \mathbb{I}\,{\otimes}\,S^z\,{-}\,S^z\,{\otimes}\,\mathbb{I})$, we hence have
\begin{align}
h_{ij}|\psi\rangle_{ij} =  S^2 \cos(Q_p) |\psi\rangle_{ij}
 \pm i S \sin(Q_p) \left( S^z_j - S^z_i \right) |\psi\rangle_{ij}.
\end{align}
This is our final result.  The $(S^z_i\,{-}\,S^z_j)$  term cancels out in the bulk  when summed over all sites (it telescopes), so 
\begin{align}
    H(\cos(Q_p))|\psi(\mathbf{Q}_p)\rangle = (S^2 \cos(Q_p) \times \# Links)|\psi(\mathbf{Q}_p)\rangle
\end{align}
as claimed.  $\blacksquare$

\newpage

\begin{thebibliography}{10}

\bibitem{Berges2004_Prethermalization}
J.~Berges, S.~Bors\'anyi, C.~Wetterich, {\it Phys. Rev. Lett.\/} {\bf 93},
  142002 (2004).

\bibitem{Rigol2007}
M.~Rigol, V.~Dunjko, V.~Yurovsky, M.~Olshanii, {\it Phys. Rev. Lett.\/} {\bf
  98}, 050405 (2007).

\bibitem{Schmiedmayer2012_Prethermalization}
M.~Gring, {\it et~al.\/}, {\it Science\/} {\bf 337}, 1318 (2012).

\bibitem{Hadzibabic2018_PrethermalBosons}
C.~Eigen, {\it et~al.\/}, {\it Nature\/} {\bf 563}, 221 (2018).

\bibitem{abanin2019}
D.~A. Abanin, E.~Altman, I.~Bloch, M.~Serbyn, {\it Rev. Mod. Phys.\/} {\bf 91},
  021001 (2019).

\bibitem{Choi2017_TimeCrystal}
S.~Choi, {\it et~al.\/}, {\it Nature\/} {\bf 543}, 221 (2017).

\bibitem{Monroe2017_TimeCrystal}
J.~Zhang, {\it et~al.\/}, {\it Nature\/} {\bf 543}, 217 (2017).

\bibitem{Arute2019}
F.~Arute, {\it et~al.\/}, {\it Nature\/} {\bf 574}, 505 (2019).

\bibitem{Flamini2019}
F.~Flamini, N.~Spagnolo, F.~Sciarrino, {\it Rep. Prog. Phys.\/} {\bf 82},
  016001 (2018).

\bibitem{JordanWigner}
P.~{Jordan}, E.~{Wigner}, {\it Zeitschrift f{\"u}r Physik\/} {\bf 47}, 631 (1928).

\bibitem{Sachdev2011}
S.~Sachdev, {\it Quantum Phase Transitions\/} (Cambridge University Press,
  2011), second edn.

\bibitem{BetheAnsatz}
H.~Bethe, {\it Zeitschrift f{\"u}r Physik\/} {\bf 71}, 205 (1931).

\bibitem{Franchini2017}
F.~Franchini, {\it An Introduction to Integrable Techniques for One-Dimensional
  Quantum Systems\/} (Springer International Publishing, 2017).

\bibitem{Znidaric2011}
M.~Žnidarič, {\it Phys. Rev. Lett.\/} {\bf 106} (2011).

\bibitem{Ljubotina2017}
M.~Ljubotina, M.~{\v Z}nidari{\v c}, T.~Prosen, {\it Nature Communications\/}
  {\bf 8}, 16117 (2017).

\bibitem{KPZ_2019_theory}
M.~Ljubotina, M.~{\v Z}nidari{\v c}, T.~Prosen, {\it
  Phys. Rev. Lett.\/} {\bf 122}, 210602 (2019).

\bibitem{bloch2021_KPZ}
D.~{Wei}, {\it et~al.\/}, arXiv:2107.00038 ([cond-mat.quant-gas] 30 June 2021).

\bibitem{popkov2021phantom}
V.~Popkov, X.~Zhang, A.~Kl\"umper, {\it Phys. Rev. B\/} {\bf 104}, L081410
  (2021).

\bibitem{Cassidy2011}
A.~C. Cassidy, C.~W. Clark, M.~Rigol, {\it Phys. Rev. Lett.\/} {\bf 106},
  140405 (2011).

\bibitem{Jepsen2020_SpinTransport}
P.~N. Jepsen, {\it et~al.\/}, {\it Nature\/} {\bf 588}, 403–407 (2020).

\bibitem{Jepsen2021_TransverseSpin}
P.~N. {Jepsen}, {\it et~al.\/}, arXiv:2103.07866 [cond-mat.quant-gas] (14 March 2021)

\bibitem{Serbyn2021}
M.~Serbyn, D.~A. Abanin, Z.~Papi{\'{c}}, {\it Nature Physics\/} {\bf 17}, 675
  (2021).

\bibitem{Svistunov2003_CounterflowSF}
A.~B. Kuklov, B.~V. Svistunov, {\it Phys. Rev. Lett.\/} {\bf 90}, 100401
  (2003).

\bibitem{Duan2003_ControllingSpinExchange}
L.-M. Duan, E.~Demler, M.~D. Lukin, {\it Phys. Rev. Lett.\/} {\bf 91}, 090402
  (2003).

\bibitem{GarciaRipoll2003_BosonsInOpticalLattice}
J.~J. Garc\'ia-Ripoll, J.~I. Cirac, {\it New Journal of Physics\/} {\bf 5}, 76
  (2003).

\bibitem{Altman2003_TwoComponentBosons}
E.~Altman, W.~Hofstetter, E.~Demler, M.~D. Lukin, {\it New Journal of
  Physics\/} {\bf 5}, 113 (2003).

\bibitem{interactionSpectroscopy}
J.~Amato-Grill, N.~Jepsen, I.~Dimitrova, W.~Lunden, W.~Ketterle, {\it Phys.
  Rev. A\/} {\bf 99}, 033612 (2019).

\bibitem{Bloch2014_SpinHelix}
S.~Hild, {\it et~al.\/}, {\it Phys. Rev. Lett.\/} {\bf 113}, 147205 (2014).

\bibitem{Thywissen2015_LG_effect}
S.~Trotzky, {\it et~al.\/}, {\it Phys. Rev. Lett.\/} {\bf 114}, 015301 (2015).

\bibitem{Kokkelmans2020_interactionSpectroscopy}
T.~Secker, J.~Amato-Grill, W.~Ketterle, S.~Kokkelmans, {\it Phys. Rev. A\/}
  {\bf 101}, 042703 (2020).

\bibitem{Lhmann2012}
D.-S. L\"{u}hmann, O.~J\"{u}rgensen, K.~Sengstock, {\it New Journal of
  Physics\/} {\bf 14}, 033021 (2012).

\bibitem{Dutta2015}
O.~Dutta, {\it et~al.\/}, {\it Reports on Progress in Physics\/} {\bf 78},
  066001 (2015).

\bibitem{Juergensen2014_DensityInducedTunneling}
O.~J\"urgensen, F.~Meinert, M.~J. Mark, H.-C. N\"agerl, D.-S. L\"uhmann, {\it
  Phys. Rev. Lett.\/} {\bf 113}, 193003 (2014).

\bibitem{campbell2006_image_clock_shift}
G.~K. Campbell, {\it et~al.\/}, {\it Science\/} {\bf 313}, 649 (2006).

\bibitem{Will2010_MultiBodyInteractions}
S.~Will, {\it et~al.\/}, {\it Nature\/} {\bf 465}, 197 (2010).

\bibitem{Baier2016}
S.~{Baier}, {\it et~al.\/}, {\it Science\/} {\bf 352}, 201 (2016).

\bibitem{Hirsch1996_offsiteInteractions}
J.~C. Amadon, J.~E. Hirsch, {\it Phys. Rev. B\/} {\bf 54}, 6364 (1996).

\bibitem{Greiner2016_FermiSpinCorrelations}
M.~F. Parsons, {\it et~al.\/}, {\it Science\/} {\bf 353}, 1253 (2016).

\bibitem{Cheuk2016_SpinCorrelations}
L.~W. Cheuk, {\it et~al.\/}, {\it Science\/} {\bf 353}, 1260 (2016).

\bibitem{Greiner2017_FermiAntiferromagnet}
A.~Mazurenko, {\it et~al.\/}, {\it Nature\/} {\bf 545}, 462 (2017).

\bibitem{Bloch2013_SingleSpin}
T.~Fukuhara, {\it et~al.\/}, {\it Nature Physics\/} {\bf 9}, 235 (2013).

\bibitem{Bloch2013_BoundMagnons}
T.~Fukuhara, {\it et~al.\/}, {\it Nature\/} {\bf 502}, 76 (2013).

\bibitem{Wang2018_BetheStrings}
Z.~Wang, {\it et~al.\/}, {\it Nature\/} {\bf 554}, 219 (2018).

\bibitem{Bera2020_BetheStringsDispersion}
A.~K. Bera, {\it et~al.\/}, {\it Nature Physics\/} {\bf 16}, 625 (2020).

\bibitem{Zwierlein2019_SpinTransport}
M.~A. Nichols, {\it et~al.\/}, {\it Science\/} {\bf 363}, 383 (2019).

%%%%%%
% new items

\bibitem{shiraishi2017}
N.~Shiraishi, {\it et~al.\/}, {\it Phys. Rev. Lett.\/} {\bf 119}, 030601 (2017).

\bibitem{mark2020}
D.~K.~Mark, {\it et~al.\/}, {\it Phys. Rev. B\/} {\bf 101}, 195131 (2020).

\bibitem{chattopadhyay2020}
S.~Chattopadhyay, {\it et~al.\/}, {\it Phys. Rev. B\/} {\bf 101}, 174308 (2020).

\bibitem{moudgalya2020}
S.~Moudgalya, {\it et~al.\/}, {\it Phys. Rev. B\/} {\bf 102}, 085140 (2020).

%% new items above

\bibitem{1_RevModPhys.83.1405}
M.~A. Cazalilla, R.~Citro, T.~Giamarchi, E.~Orignac, M.~Rigol, {\it Rev. Mod.
  Phys.\/} {\bf 83}, 1405 (2011).

\bibitem{matsuda1970}
H.~Matsuda, T.~Tsuneto, {\it Progress of Theoretical Physics Supplement\/} {\bf
  46}, 411 (1970).

\bibitem{Dooley2021}
S.~Dooley, {\it PRX Quantum\/} {\bf 2}, 020330 (2021).

\bibitem{rydberg_XXZ_scholl2021}
P.~{Scholl}, {\it et~al.\/}, arXiv:2107.14459 [quant-ph] (30 July 2021).

\end{thebibliography}
\end{document}